\documentclass[fleqn,usenatbib]{mnras}
\usepackage{xspace}
\usepackage{newtxtext,newtxmath}
\usepackage[T1]{fontenc}

\DeclareRobustCommand{\VAN}[3]{#2}
\let\VANthebibliography\thebibliography
\def\thebibliography{\DeclareRobustCommand{\VAN}[3]{##3}\VANthebibliography}

\usepackage{graphicx}	% Including figure files
\usepackage{natbib}
\usepackage{academicons}
\usepackage{hyperref}
\usepackage{xcolor}
\usepackage{enumitem}
\usepackage{graphicx}	% Including figure files
\usepackage{amsmath}	% Advanced maths commands
\usepackage{orcidlink}

\newcommand{\orcid}[1]{\href{https://orcid.org/#1}{\includegraphics[width=8pt]{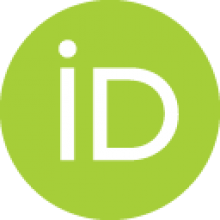}}}

\newcommand{\afe}{$\rm [\alpha/Fe]$}
\newcommand{\feh}{$\rm [Fe/H]$}
\newcommand{\logirlum}{$\log(L_{\rm IR}/L_\odot)$}
\newcommand{\logmstar}{$\log(M_{\star}/M_\odot)$}

\newcommand{\galaxyname}{RUBIES-EGS-QG-1}
\newcommand{\alphamc}{$\alpha$-MC}

\usepackage[normalem]{ulem}
\title[Revisiting a maximally-old galaxy at $z=4.9$]{Extreme, but not impossible: Self-consistent stellar population models alleviate the implausibly efficient formation of a massive quiescent galaxy at $z=4.9$}

\newcommand{\nhfpthanks}{\thanks{NHFP Fellow}}

\author[Park et al.]
{Minjung Park$^{\orcidlink{0000-0002-8435-9402}1, 2}$\thanks{E-mail: mp2246@cam.ac.uk},
David J. Setton$^{3}$\nhfpthanks,
Anna de Graaff$^{4,5}$\thanks{Clay Fellow},
Joel Leja$^{6,7,8}$,
Leindert A. Boogaard$^{9}$,
\newauthor 
Sandro Tacchella$^{\orcidlink{0000-0002-8224-4505}1,2}$,
Benjamin D. Johnson$^{4}$,
Charlie Conroy$^{4}$,
Rachel Bezanson$^{10}$,
Andy D. Goulding$^{11}$,
\newauthor
Jenny Greene$^{11}$,
Michaela Hirschmann$^{12}$,
Michael V. Maseda$^{13}$
Ian McConachie$^{\orcidlink{0000-0002-2446-8770}14}$,
Rohan Naidu$^{15}$,
\newauthor
Francesco Valentino$^{16,17}$,
Fabian Walter$^{5}$,
Bingjie Wang$^{\orcidlink{0000-0001-9269-5046}11}${\color{blue}{\footnotemark[2]}},
Katherine E. Whitaker$^{16,18}$,
Yunchong Zhang$^{\orcidlink{0000-0001-6454-1699}10}$
\\
\\
$^{1}$Kavli Institute for Cosmology, University of Cambridge, Madingley Road, Cambridge CB3 0HA, UK\\
$^{2}$Cavendish Laboratory, University of Cambridge, 19 JJ Thomson Avenue, Cambridge CB3 0HE, UK\\
$^{3}$The William H. Miller III Department of Physics and Astronomy, Johns Hopkins University, 3400 N. Charles Street, Baltimore, MD 21218\\
$^{4}$Center for Astrophysics | Harvard \& Smithsonian, 60 Garden Street, Cambridge, MA 02138, USA \\
$^{5}$Max-Planck-Institut für Astronomie, Königstuhl 17, D-69117 Heidelberg, Germany\\
$^{6}$Department of Astronomy \& Astrophysics, The Pennsylvania State University, University Park, PA 16802, USA\\
$^{7}$Institute for Computational \& Data Sciences, The Pennsylvania State University, University Park, PA 16802, USA\\
$^{8}$Institute for Gravitation and the Cosmos, The Pennsylvania State University, University Park, PA 16802, USA\\
$^{9}$Leiden Observatory, Leiden University, PO Box 9513, 2300 RA Leiden, The Netherlands\\
$^{10}$Department of Physics and Astronomy and PITT PACC, University of Pittsburgh, Pittsburgh, PA 15260, USA\\
$^{11}$Department of Astrophysical Sciences, Princeton University, 4 Ivy Lane, Princeton, NJ 08544, USA\\
$^{12}$Institute of Physics, Laboratory of Astrophysics, EPFL, Observatoire de Sauverny, Chemin Pegasi 51, 1290 Versoix, Switzerland\\
$^{13}$Department of Astronomy, University of Wisconsin-Madison, 475 N. Charter St., Madison, WI 53706 USA\\ 
$^{14}$Observatories of the Carnegie Institution for Science, 813 Santa Barbara Street, Pasadena, CA 91101, USA\\ 
$^{15}$Institute for Astronomy, University of Hawai’i, Honolulu, HI, USA\\
$^{16}$Cosmic Dawn Center (DAWN), Denmark\\
$^{17}$DTU Space, Technical University of Denmark, Elektrovej 327, DK-2800 Kgs. Lyngby, Denmark\\
$^{18}$Department of Astronomy, University of Massachusetts Amherst, Amherst, MA 01003, USA\\}

\date{Accepted XXX. Received YYY; in original form ZZZ}

\pubyear{\the\year{}}

\begin{document}
\label{firstpage}
\pagerange{\pageref{firstpage}--\pageref{lastpage}}
\maketitle

\begin{abstract}
Recent JWST spectroscopy has identified ``maximally old'' massive quiescent galaxies at $z>3$ whose inferred formation histories suggest that >100\% of the available baryons have been converted into stars within only a few hundred Myr.
\galaxyname\ is a massive ($\log(M_\star/M_\odot)\sim11.0$) quiescent galaxy at $z_{\rm spec}=4.9$ that may require such impossibly efficient formation. 
In this work, we reanalyse \galaxyname\ incorporating new constraints from MIRI and NOEMA observations and new self-consistent stellar population synthesis (SPS) models with non-solar \afe\ covering the full wavelength range. 
We find that fitting the rest-UV continuum with previous inconsistent spectral models can incorrectly infer a maximally old stellar age. 
The new self-consistent models instead infer that the galaxy experienced an intense starburst $\approx300\,\rm Myr$ before observation, followed by rapid quenching, resulting in a young ($\sim0.3\,\rm Gyr$), $\alpha$-enhanced (\afe$\sim0.4$), metal-rich ($\log(Z/Z_\odot)\approx-0.1$) stellar population. 
This formation history, with a lower stellar age (by $\sim0.3\,\rm Gyr$) and mass (by $\sim0.2\,\rm dex$), is consistent with a much lower efficiency of $\sim30\%$, although still requiring a high peak efficiency of $\sim60\%$ during the starburst.
The self-consistent models also predict an infrared luminosity of $L_{\rm IR}\approx10^{12}\,L_\odot$, suggesting that the galaxy remains IR-luminous after quenching because its exceptionally compact stellar population, formed in the recent starburst, continues to heat the dust.  
This work demonstrates the importance of physically self-consistent SPS models for accurately inferring the physical properties of early massive galaxies. 
Future work applying such models to larger samples will reveal how they shift the maximally old population.

\end{abstract}


%%%%%%%%%%%%%%%%%%%%%%%%%%%%%%%%%%%%%%%%%%%%%%%%%%

%%%%%%%%%%%%%%%%% BODY OF PAPER %%%%%%%%%%%%%%%%%%
\clearpage

\vspace{-0.5cm}
\section{Introduction}
\label{sec:intro}
Recent deep JWST observations have revealed a surprisingly abundant population of massive ($>10^{10}\,M_\odot$) quiescent galaxies at $z>3$ \citep[e.g.,][]{Valentino2023ApJ, Long2024CEERS, Carnall2024excels, Nanayakkara2024, Baker2024, Ito2025_deepdive, Zhang2026}. 
Some of these early massive quiescent galaxies exhibit puzzling formation histories, appearing too old and too massive, often referred to as ``maximally-old'' galaxies \citep[e.g.,][]{Glazebrook2024, Carnall2024excels, deGraaff2025_z4.9, Turner2025, Weibel2025_z7.3, McConachie2026}. 
Their inferred star formation histories (SFHs) indicate that they assemble $M_\star\approx10^{11}\,M_\odot$ at very early epochs, as early as $z\approx10$, by intense starbursts.  
Immediately after this explosive formation, the star formation is shut down by $z>7$. 
This rapid and early assembly of high stellar mass requires exceptionally efficient conversion of baryons into stars within only a few hundred Myr, with inferred baryon-to-star conversion efficiencies exceeding 100\%.
Such highly efficient formation is not reproduced in any of the current state-of-the-art large-volume cosmological simulations \citep[e.g., ][]{Kannan2022, Kannan2023, Schaye2023, Lovell2023_flare_sim, Kimmig2023, Lagos2025}, challenging our understanding of galaxy formation models.

Several theoretical mechanisms have been proposed to understand unusually high star-formation efficiencies at cosmic dawn. 
These include rapid starbursts in dense gas that convert a substantial fraction of the gas into stars before supernova feedback can respond (i.e., ``feedback-free'' starbursts), the breakdown of feedback regulation at very high gas surface densities, and an increasing fraction of gas residing in dense star-forming clouds at high redshifts \citep[e.g.,][]{Grudic2018, Dekel2023, Li2024, Somerville2025}. 
Although these theoretical models differ in detail, they all point to the idea that the dense, rapidly accreting environments of the early Universe may enable gas to be converted into stars more efficiently than in typical lower-redshift galaxies.

\galaxyname\ is one of the earliest spectroscopically-confirmed massive ($\log(M_\star/M_\odot)\approx11.0$) quiescent galaxies at $z=4.9$, reported in \cite{deGraaff2025_z4.9}.  
Based on spectral energy density (SED) fitting to both NIRCam photometry and NIRSpec/PRISM spectroscopic data, they have identified two viable SFH solutions, with the fiducial solution placing the galaxy among the maximally-old population. 
In this fiducial solution, it assembles most of its stellar mass at $z\approx10.0$ with an intense starburst ($\Delta t\approx180\,\rm Myr$), followed by rapid quenching at $z\approx8.6$. 
This formation history requires extremely high, potentially unphysical, baryon-to-star conversion efficiency exceeding 100\%. 
The fiducial SED fit also favours a low stellar metallicity ($Z\approx0.2\,Z_\odot$) and low dust attenuation (with $V$-band attenuation of $A_V\approx0.17$ mag). 
However, the adopted stellar population models assume solar-scaled chemical abundance patterns, which may lead to incorrect stellar metallicity and other stellar population properties. 
When the stellar metallicity is instead fixed to $Z=Z_\odot$, the fit favours a much younger stellar population and higher dust attenuation (with $A_V\approx 0.48$ mag). 
The corresponding mass assembly history requires a much lower baryon-to-star conversion efficiency ($\sim20\%$), alleviating the apparent tension with galaxy formation models.
The two viable solutions illustrate the well-known degeneracy between age, dust attenuation, and metallicity. 
In summary, these results demonstrate that the physical interpretation of this galaxy remains uncertain and strongly depends on the adopted stellar population models, motivating a re-analysis with additional observational constraints and improved stellar population models.

One way to help break this degeneracy is through IR and sub-mm observations, which provide additional constraints on the level of dust emission in stellar population modelling.  
The 8-band NIRCam photometry used in the original analysis by \cite{deGraaff2025_z4.9} spans only the rest-frame UV to optical wavelengths and therefore cannot constrain the SED shape beyond $\lambda_{\rm rest}>0.8\,\rm um$. 
Because the two viable solutions (old vs. young) identified in the original analysis suggest different amounts of dust attenuation ($A_V=0.17 $ vs. $A_V=0.48$), they also predict different levels of IR emission through energy balance \citep[e.g.,][]{daCunha2008, Noll2009, Leja2017, Salim2018, Boquien2019, Johnson2021}.
Incorporating IR data into the SED fitting can therefore distinguish between the two solutions by breaking the age–dust–metallicity degeneracy, leading to more robust estimates of the stellar age (and as a result, metallicity).

An additional approach to better constraining the stellar population of these high-$z$ massive quiescent galaxies is to employ stellar population models that incorporate non-solar chemical abundance patterns. 
In particular, high-redshift galaxies are expected to have non-solar abundance ratios between $\alpha$ elements (e.g., O, Mg, Si, and Ca) and Fe-peak elements, i.e., \afe. 
These two elemental groups are enriched by different sources on distinct timescales \citep[e.g.,][]{Nomoto1984, Woosley1995, Nomoto2013, Kobayashi2020}: $\alpha$ elements are enriched on a short timescale ($\sim 20\,\rm Myr$) by core-collapse supernovae (SNe II), while Fe-peak elements are enriched both by SNe II and SNe Ia on a much longer timescale ($\sim1\,\rm Gyr$). 
In the young Universe, where there was less time for SNe Ia, chemical enrichment was primarily driven by SNe II; thus, high-$z$ galaxies are likely $\alpha$-enhanced (with high \afe). 
Several studies have shown that massive quiescent galaxies at high redshifts are $\alpha$-enhanced \citep[e.g.,][]{Kriek2016Nature, Jafariyazani2020ApJ, Beverage2024_suspense, Hamadouche2026, Leung2026, Edward2026}. 
Using self-consistent $\alpha$-enhanced stellar population models, a recent study of \cite{Park2025} has shown that $\alpha$-enhancement has significant impacts on broadband colors (up to $\sim0.2\,\rm mag$ with \afe$=+0.4$) and individual spectral features sensitive to $\alpha$-elements \citep[see also][]{Coelho2007, Knowles2023}. 
Therefore, to infer accurate physical properties of high-$z$ quiescent galaxies, it is essential to adopt self-consistent stellar population models that incorporate non-solar \afe.

In this paper, we re-analyse \galaxyname\ incorporating 1) new constraints from MIRI and NOEMA observations and 2) novel, self-consistent $\alpha$-enhanced stellar population models to obtain more robust constraints on its SFH and the corresponding baryon-to-star conversion efficiency and to assess whether the inferred efficiency can be reduced to a physically plausible value.
We describe the photometric and spectroscopic observations and spectral energy distribution (SED) fitting in Section~\ref{sec:data_and_sed_fitting}. 
In Section~\ref{sec:result}, we present our fitting results with different SPS models and inferred SFHs.
In Section~\ref{sec:discussion}, we discuss the stellar mass assembly history and the inferred dust and IR luminosity properties.  
Finally, we present a summary and our conclusion in Section~\ref{sec:summary_conclusion}.
Throughout the paper, we adopt the Chabrier initial mass function (IMF) \citep{Chabrier2005The2005}, as in the original analysis of \cite{deGraaff2025_z4.9}. We adopt the WMAP9 cosmology \citep{Hinshaw2013ApJS_WMAP9} unless noted otherwise. 

\vspace{2cm}

\section{Data and SED fitting}
\label{sec:data_and_sed_fitting}

\subsection{Photometry}
\begin{figure*}
    \centering
    \includegraphics[width=\textwidth]{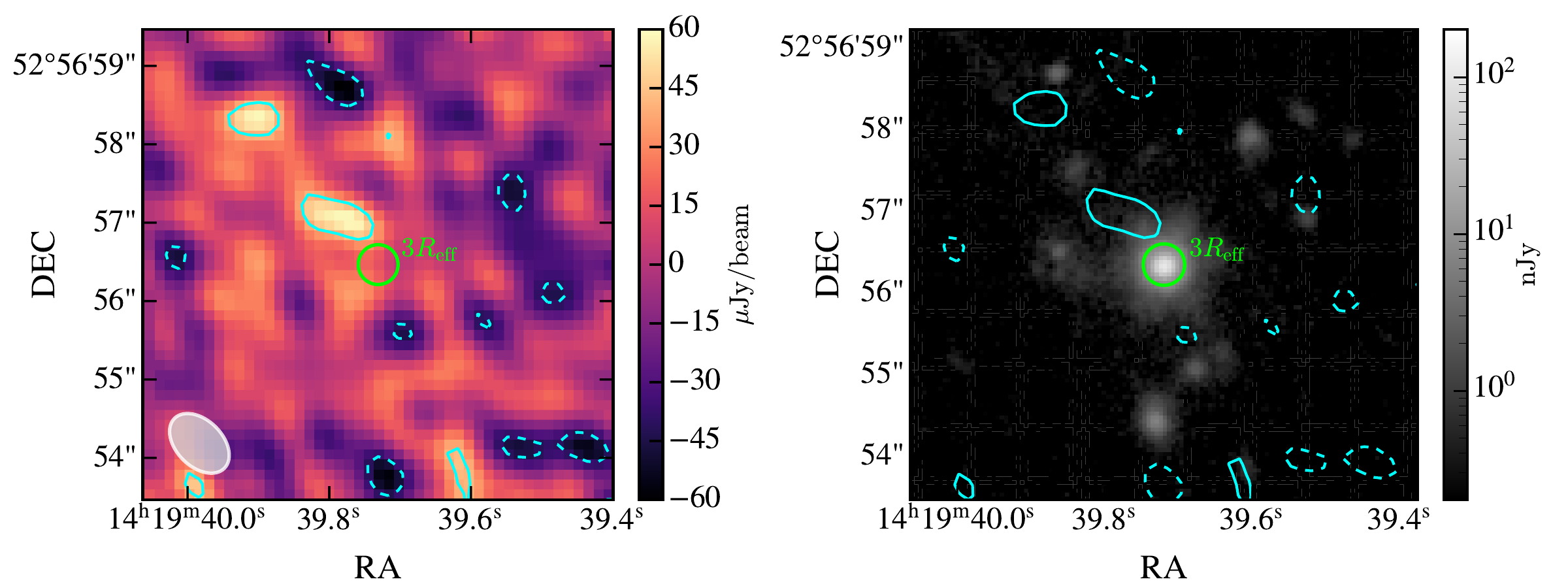}
    \caption{(Left) NOEMA continuum image in $6''\times6''$ cutout. The solid and dashed cyan contours indicate $+2/-2\,\rm \sigma$ detections, respectively. The beam size is shown as the white ellipse at the lower left corner. 
    The green circle indicates three times the rest-optical effective radius of \galaxyname\ ($R_{\rm eff}=0.55\,\rm kpc$).
    The \galaxyname\ source is not detected in NOEMA; the target reaches $20.3\,\rm \mu Jy/beam$ RMS in the continuum at 253 GHz. 
    (Right) NIRCam/F444W image in the same $6''\times6''$ cutout. Same as in the left panel, $+2/-2\,\rm \sigma$ NOEMA contours are shown as solid and dashed cyan lines, respectively, and the green circle indicates a radius of $3\,R_{\rm eff}$.}
    \label{fig:noema_map}
\end{figure*}

We collect NIRCam imaging in 8 bands (F090W, F115W, F150W, F200W, F277W, F356W, F410M, and F444W), combined from the publicly available CEERS program (ERS-1345, PI: S Finkelstein, \citealt{Bagley2023}) and GO-2234 program (PI: Banados; Khusanova et al. in prep). 
We use the same NIRCam data reduction as in the original analysis of \cite{deGraaff2025_z4.9}.
Using the reduced image mosaics from the DAWN JWST Archive (DJA; version 7.4) and empirical point spread function models (PSF), we construct mosaics PSF-matched to the F444W mosaic \citep[see][for details]{Weibel2024}. 
We use circular apertures with a radius of 0.25 arcsec to measure fluxes from the PSF-matched mosaics. 
The fluxes are then corrected to total fluxes, as described in \cite{Weibel2025_z7.3}.
See \cite{deGraaff2025_z4.9} for the details of the NIRCam photometric data reduction.

We obtain 4 band MIRI photometry (F770W, F1000W, F1500W, and F2100W) by combining publicly available data from program MEGA \citep[GO- 3794, PI: A. Kirkpatrick,][]{Backhaus2025_MEGA}. Using the reduced image mosaics from the DAWN JWST Archive (DJA; version 7.4), we obtain total fluxes and uncertainties by measuring the photometry in apertures and correcting to total using the enclosed-energy corrections for these filters defined in \cite{Alberts2024_SMILES}. All four bands are detected, with total fluxes of $4.21\pm0.05$, $4.44\pm0.08$, $2.11\pm0.18$, and $2.04\pm0.47$ $\mu$Jy in F770W, F1000W, F1500W, and F2100W, respectively.

Finally, we obtain unresolved NOEMA observations at 253\,GHz (1.2\,mm) as part of program W24EE (PIs: M. Maseda, A. de Graaf). 
A single good-quality track was observed on Jan 13, 2025, which was calibrated and imaged in \textsc{gildas}.  
The bandpass calibrator was 0851+202, the flux calibrator was LkHa101, and the phase calibrators were J1439+499 and 1418+546. 
We inspect the data for parasites and frequency ranges with elevated noise and mask channels accordingly before creating the continuum map.  
An initial inspection yields no significant emission at the position of \galaxyname, but a bright source is present at the edge of the field, and we \textsc{hogbom} clean the image down to 1\,$\sigma$ using a mask around this source.  
The final cleaned continuum map has a synthesised natural beam size of $0\farcs90\times0\farcs59$ (PA=45\, degrees) and reaches a sensitivity of 20.3\,$\mu$Jy\,beam$^{-1}$. 
Fig.~\ref{fig:noema_map} shows the NOEMA continuum image in a $6''\times6''$ cutout. 
The beam size is shown as the white ellipse at the lower left corner.
The solid and dashed cyan contours indicate $+2/-2\,\rm \sigma$ detections, respectively. 
The green circle indicates three times the rest-optical effective radius ($R_{\rm eff}$) of \galaxyname\ ($R_{\rm eff}=0.55\,\rm kpc$).
\galaxyname\ is not detected in NOEMA, which, given the source is not resolved, implies a $20.3\,\rm \mu Jy\,beam^{-1}$ upper limit ($1\,\sigma$) on the continuum flux density at 253 GHz.

\subsection{Spectroscopy}
The NIRSpec/PRISM spectroscopic data are obtained from the RUBIES program (Cycle 2, GO-4233; PIs: A. de Graaff and G. Brammer, \citealt{deGraaff2025_rubies}), observing galaxies in the CANDELS EGS and UDS deep fields using the NIRSpec microshutter array (MSA). 
The MSA pointings were observed for 48 minutes with PRISM/CLEAR using 3 microshutters and a 3-point nodding pattern. 
We use the same NIRSpec data reduction as in \cite{deGraaff2025_z4.9}.
The NIRSpec data are reduced using the \texttt{msaexp} pipeline \citep{Brammer2023} version 3, as described in detail in \cite{Heintz2025} and \cite{deGraaff2025_rubies}. 
We use empty sky shutters to derive the bar shadow corrections and construct the master background. 
We scale up the observed $1\,\rm \sigma$ uncertainties on the 1D extracted spectrum by a factor of 1.7 to account for underestimated uncertainties given the pixel-to-pixel variations. 
The final flux calibration of the 1D spectrum is performed by matching the continuum level to the multi-band photometry from JWST/NIRCam, JWST/MIRI, and NOEMA through SED fitting, as described in the following section.

\subsection{Prospector SED fitting}
\begin{table*}
\centering
\caption{Summary of \texttt{Prospector} model parameters and priors used.}
\label{tab:prospector_params}
\begin{tabular*}{\textwidth}{@{\extracolsep{\fill}}lllll}
\hline
- & Parameter & Unit & Prior& Note \\
\hline

\multicolumn{4}{l}{\textit{Stellar properties}} \\
& Redshift & $z_{\rm spec}$ & Flat prior, $[z_{\rm spec} - 0.005,\, z_{\rm spec} + 0.005]$ & $z_{\rm spec}=4.9086$ \\
& Stellar velocity dispersion & $\sigma_\star\, [\mathrm{km\ s^{-1}}]$ & Flat prior, [50, 2000] & NIRSpec/PRISM R$\sim100$ \\
& Total stellar mass formed & $\log(M_\ast\ /\ M_\odot)$ & Flat prior [8.5, 13.0] & - \\
& Total stellar metallicity & $\log(Z/\ Z_\odot)$ & Flat prior, [-2.0, +0.5] 
& Default models \\
& \feh & - & Flat prior, [-2.0, +0.5]
& \alphamc\ models \\
& \afe & - & Flat prior, [-0.2, +0.6]
& \alphamc\ models \\
& SFH & $\Delta \log\rm SFR$ & Student's T-distribution ($\mu=0, \sigma=0.3, \nu=2.0$) & \cite{Leja2019HowNonparametric} \\ 

\hline
\multicolumn{4}{l}{\textit{Dust attenuation and emission}} \\
& Optical depth in $V$-band & $\tau_{\rm old}$ & Flat prior [0.0, 3.68] & - \\
& Extra attenuation toward young stars & $\tau_{\rm young}$ & $\tau_{\rm young}$/$\tau_{\rm old} \sim N(\mu=1, \sigma=0.3)$ with min$=0.0$, max$=2.0$& - \\
& A power law modifier to Calzetti slope & $n_{\rm dust\,index}$ & $N(\mu=0.0, \sigma=0.3)$ with min$=-1.5$, max$=4.0$ & \cite{Kriek_Conroy_dust_2013} \\
& PAH dust mass fraction & $q_{\rm PAH} (\%)$ & Flat prior [0.5, 10.0] & \cite{Draine_Li_2007} \\
& Contribution of dust heated with $U>U_{\rm min}$ & $\gamma$ & Flat prior, [0.0, 1.0] & \cite{Draine_Li_2007} \\
& Minimum radiation field & $U_{\rm min}$ & Flat prior [0.1, 25.0] & \cite{Draine_Li_2007} \\

\hline
\end{tabular*}
\end{table*}

We use \texttt{Prospector} \citep{Johnson2021}, a fully Bayesian stellar population inference code, to fit both observed photometry and NIRSpec/PRISM data. 
For observed photometry, we use all the available photometry (NIRCam + MIRI + NOEMA), unless noted otherwise. 
We cap the signal-to-noise ratio ($\rm S/N$) of all photometric measurements at $\rm S/N=20$, as in the original analysis of \cite{deGraaff2025_z4.9}, to account for any observational systematics.  
\texttt{Prospector} models the stellar population properties of galaxies with a set of parameters, including redshift ($z_{\rm redshift}$), stellar mass ($\log(M_\star/M_\odot)$), stellar velocity dispersion ($\sigma_\star$), stellar metallicity, and a non-parametric SFH. 
We use 14 SFH time bins, starting with the two youngest bins of [0, 10 Myr] and [10 Myr, 50 Myr]. 
The time bins are then equally spaced logarithmically until the age of the Universe. 
The choice of SFH time bins does not affect our results.
We use the continuity SFH prior \citep{Leja2019HowNonparametric}, which fits the logarithmic ratio between the star formation rates (SFRs) in the two adjacent time bins with a Student’s T-distribution prior centred at $\mu=0$ with $\sigma=0.3$ and $\nu=2.0$.
The impact of the choice of the SFH priors is explored in the original analysis. 
To more clearly assess the impact of the IR data and the adopted stellar population models, we stick to the same continuity SFH priors adopted in the original analysis in \cite{deGraaff2025_z4.9}.

We use two stellar population synthesis (SPS) models, synthesised using FSPS \citep{Conroy2009_FSPS, Conroy2010_FSPS2}:

\begin{enumerate}[leftmargin=2em, itemsep=0.7em]
    \item Default: MIST v1.2 isochrones + MILES spectral library
    \\ \hspace*{4cm} (C3K v1.3 outside rest-optical)
    \item \alphamc: MIST v2.5 isochrones + C3K v2.3 spectral library
\end{enumerate} 
The default models are the SPS models adopted in the original analysis in \cite{deGraaff2025_z4.9}, which use the solar-scaled MIST v1.2 isochrones \citep{Choi2016} and the MILES stellar spectral library \citep{Sanchez-Blazquez2006}. 
Note that MILES covers only the rest-optical (3525-7500\text{\AA}), so in FSPS, it is stitched to the C3K v1.3 stellar library \citep{Cargile2020_C3K} outside the rest-optical wavelengths.
The \alphamc\ models are self-consistent $\alpha$-enhanced SPS models, constructed from $\alpha$-enhanced MIST v2.5 isochrones \citep{Dotter2026} and an updated, $\alpha$-enhanced C3K v2.3 spectral library. 
The \alphamc\ models have a wide parameter coverage in terms of age ($\rm \log(age/yr=5.0-10.3)$) and metallicity (\feh$=[-2.5, +0.5]$ in steps of 0.25 and \afe$=[-0.2, +0.6]$ in steps of 0.2) and self-consistently cover wide wavelength range ($\rm 0.1-2.5\,\mu m$). 
See \cite{Park2025} for further details of the models. 
For both SPS models, the Chabrier IMF \citep{Chabrier2005The2005} is assumed as in the original analysis of \cite{deGraaff2025_z4.9}.

We assume the two-component dust model (birth-cloud component and diffuse dust component), following \cite{CharlotFall2000}. 
We model dust attenuation with three parameters: $\tau_{\rm old}$ (optical depth in V band; $A_V=1.086\,\tau_{\rm old}$), $\tau_{\rm young}$ (extra attenuation toward young stars, and $n_{\rm dust\,index}$ (a power-law modifier to the \cite{Calzetti2000} law, following \cite{Kriek_Conroy_dust_2013}). 
Dust emission is modelled using three parameters ($q_{\rm PAH}$, $\gamma$, and $U_{\rm min}$) following \cite{Draine_Li_2007}. 
The priors used for the model parameters are summarised in Table.~\ref{tab:prospector_params}.
We assume that the total amount of energy absorbed by dust is re-emitted in the IR.
We did not include AGN emission in our models. 
We also test SED fitting with AGN emission models included and confirm that including AGN emission models does not affect our results or conclusions.

For the NIRCam/PRISM spectroscopic data, we fit $1200-8000$ \text{\AA} in the rest frame.  
We use one ``jitter'' parameter (\texttt{spec\_jitter}) for the spectrum fitting that rescales the observed spectrum uncertainties to obtain a good fit. 
We also introduce one parameter for the spectral fitting that characterises the fraction of outliers in the spectral pixels \citep[see Appendix D in][for the outlier model]{Johnson2021}.
To account for the flux calibration and slit-loss corrections, we adopt a polynomial distortion of the spectrum to match the observed spectral/continuum shape to that of the SED model. 
We use a polynomial order of 6, as in the original analysis of \citep{deGraaff2025_z4.9}.
We have tested different polynomial orders and confirmed that our main results are robust to the adopted polynomial order.
We also account for the nebular emission lines, including Ly$\alpha$, [O II], [O III], H$\delta$, H$\gamma$, H$\beta$, H$\alpha$, [N II], and [S II] using the emission-line marginalisation models in \texttt{Prospector}.

\section{Result}
\label{sec:result}

\subsection{SED fitting with additional IR constraints and new $\alpha$-enhanced SPS models}
\label{sec:fitting_new_data_new_models}

\begin{figure*}
    \centering
    \includegraphics[width=0.93\textwidth]{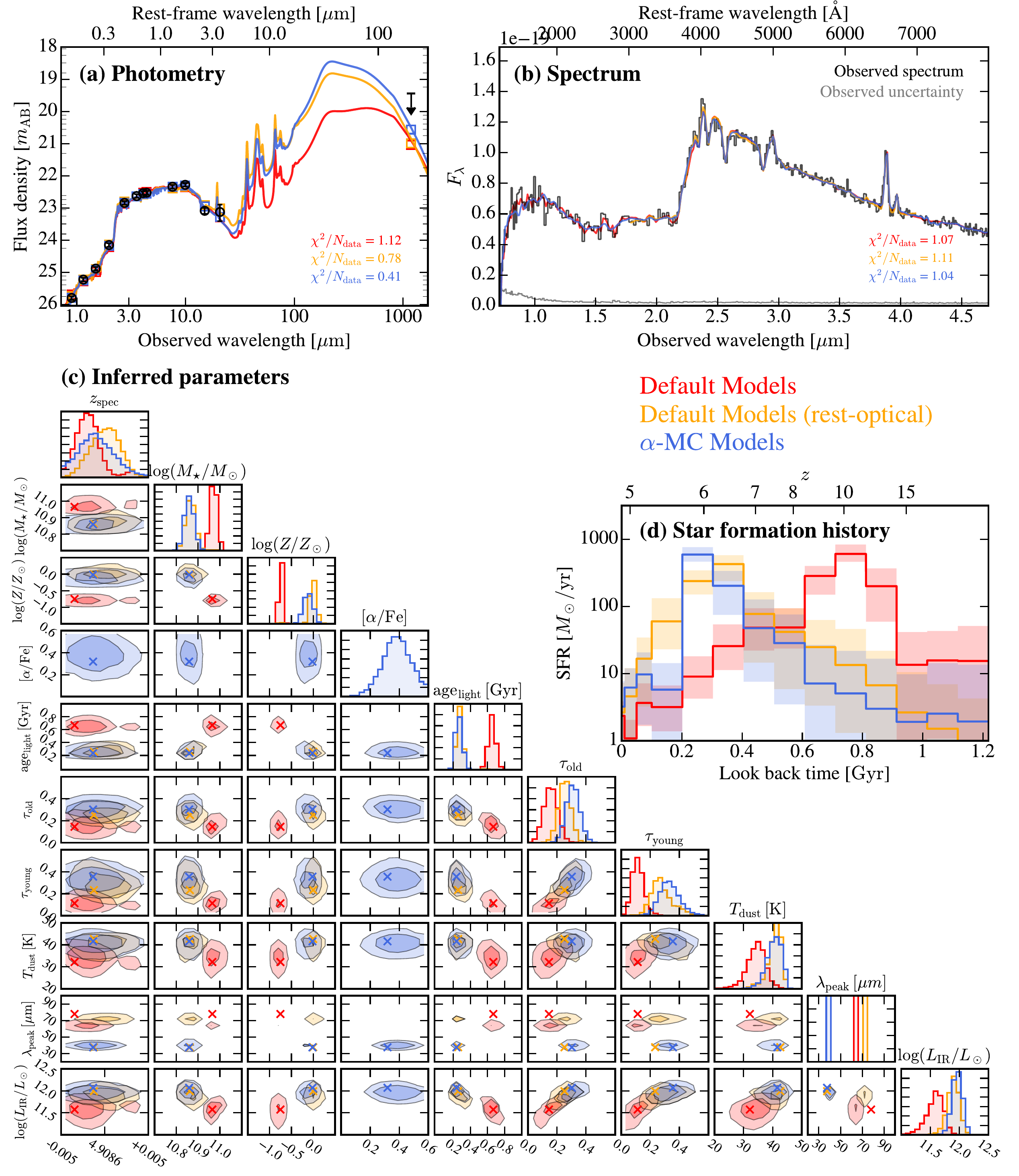}
    \caption{Prospector SED-fitting results with the updated photometry (including new MIRI and NOEMA observations) using different SPS models. 
    (a) Fit to photometry. Observed 8-band NIRCam photometry and 4-band MIRI photometry are shown as black circles with uncertainties shown as error bars. 
    The $3\,\rm \sigma$ upper limit of the NOEMA detection is shown as a black arrow.
    The best-fit SED models are shown as solid lines.
    (b) Fit to the spectrum. The observed spectrum and its uncertainties are shown in black and grey, respectively, and the best-fit spectra are shown as solid lines. 
    (c) Corner plot illustrating the joint distributions and covariances between model parameters.
    The contours in the off-diagonal panels indicate $1\,\rm \sigma$ and $2\,\rm \sigma$ of the joint posterior distributions, and the crosses indicate the MAP values.
    (d) Nonparametric SFH inferred using the continuity SFH prior. The solid line shows the 50th percentile of the posteriors in each time bin, with the shaded region indicating $1\,\sigma$.
    Colours (lines and contours) in all panels indicate the SPS models adopted: 
    \textbf{Red}: the default models used in the original analysis. 
    \textbf{Yellow}: the same default models but fitting the NIRSpec/PRISM data only over the rest-frame optical range, $0.3-0.7\,\rm \mu m$.  
    \textbf{Blue}: \alphamc\ models with \afe\ as an additional free parameter. 
    We find that the two SPS models (default vs. \alphamc) infer two distinct solutions for this galaxy. 
    The default models favour an old ($\sim0.6,\rm Gyr$), metal-poor, and dust-poor population, whereas the \alphamc\ models favour a younger ($\sim0.3,\rm Gyr$), $\alpha$-enhanced (\afe$\sim0.4$), near-solar-metallicity population with higher dust attenuation.
    Interestingly, restricting the default models to fit the NIRSpec/PRISM data only over the rest-optical range ($0.3-0.7\,\rm \mu m$), the native coverage of the MILES library, also favours the young solution (yellow).
    This suggests that the old solution inferred by the default SPS models is primarily driven by fitting the rest-UV continuum with inconsistent spectral models (i.e., MILES and C3K are stitched together across wavelength ranges in the default models) and highlights the importance of using SPS models that self-consistently cover the full wavelength range. 
    } 
    \label{fig:allphot_allmodels}
\end{figure*}

\begin{table*}
\setlength{\tabcolsep}{5pt}
\centering
\caption{Best-fit model parameters inferred using different SPS models: 
redshift ($z_{\rm spec}$), final (survived) stellar mass ($\log(M_\star/M_\odot)$), total stellar metallicity ($\log(Z/Z_\odot)$), \afe, light-weighted stellar age, $\tau_{\rm old}$, $\tau_{\rm young}$, dust temperature ($T_{\rm dust}$), and IR luminosity (\logirlum).
The values correspond to the median of the posterior distributions, with uncertainties indicating the 16th-84th percentile range.}
\label{tab:inferred_params}
\begin{tabular}{lccccccccc}
\hline
SPS models (different colours in Fig.~\ref{fig:allphot_allmodels})&
$z_{\rm spec}$ &
$\log(M_\star/M_\odot)$ &
$\log(Z/Z_\odot)^{\dagger}$ &
\afe\ &
age$_{\rm light}$ &
%$A_V$ &
$\tau_{\rm old}$ &
$\tau_{\rm young}$ &
$T_{\rm dust}$ &
$\log(L_{\rm IR}/L_\odot)$ \\
&
&
&
&
&
[Gyr] &
&
&
[K] &
\\
\hline
Default models (red)
& 4.0967 & $10.97^{+0.01}_{-0.02}$ & $-0.75^{+0.03}_{-0.02}$ & -- & $0.65^{+0.04}_{-0.04}$ & $0.16^{+0.05}_{-0.05}$ & $0.11^{+0.04}_{-0.04}$ & $34^{+3}_{-3}$ & $11.6^{+0.1}_{-0.2}$ \\

Default models (optical only, yellow)
& 4.0987 & $10.87^{+0.02}_{-0.02}$ & $-0.00^{+0.07}_{-0.06}$ & -- & $0.27^{+0.04}_{-0.03}$ & $0.25^{+0.04}_{-0.04}$ & $0.27^{+0.09}_{-0.06}$ & $41^{+2}_{-3}$ & $11.9^{+0.1}_{-0.1}$   \\

\alphamc\ models (blue)
& 4.0975 & $10.86^{+0.02}_{-0.02}$ & $-0.12^{+0.10}_{-0.11}$ & $0.38^{+0.10}_{-0.10}$ & $0.26^{+0.04}_{-0.04}$ & $0.31^{+0.05}_{-0.05}$ & $0.33^{+0.07}_{-0.06}$ & $41^{+2}_{-3}$ & $12.0^{+0.1}_{-0.1}$  \\
\hline
\end{tabular}
\begin{minipage}{\textwidth}

\footnotesize

$^{\dagger}$ Note that while MIST v1.2 isochrones used in Default models adopt solar abundance ($Z_\odot$) from \cite{Asplund2009}, \alphamc\ models adopt the solar abundance from \cite{GS98} consistently for both isochrones and stellar spectral library.

\end{minipage}
\end{table*}

We fit the updated photometry, including the new MIRI and NOEMA observations, together with the NIRSpec/PRISM spectrum using the two SPS models: Default and \alphamc. 
Fig.~\ref{fig:allphot_allmodels} shows the resulting SED fits and inferred parameters. 
Panel (a) shows the NIRCam, MIRI, and NOEMA photometric data used for fitting, displayed as black circles with error bars indicating the uncertainties. 
The $3\,\rm \sigma$ upper limit of the NOEMA detection is shown as a black arrow.
The best-fit SED model is shown as coloured solid lines. 
Throughout the figure, different colours indicate the results obtained using different SPS models (red: Default models; blue: \alphamc\ models). 
Panel (b) shows the NIRSpec/PRISM spectrum in black, with the uncertainties shown in grey. 
The best-fit spectra are overplotted as coloured solid lines.  
Panel (c) shows a corner plot of the posterior distributions for some of the model parameters. 
The diagonal panels display the marginalized/one-dimensional posterior distributions of the parameters: redshift ($z_{\rm spec}$), final (survived) stellar mass ($\log(M_\star/M_\odot)$), total stellar metallicity ($\log(Z/Z_\odot)$), \afe, light-weighted stellar age, $\tau_{\rm old}$, $\tau_{\rm young}$, dust temperature ($T_{\rm dust}$), the wavelength of the peak of the infrared/dust SED ($\lambda_{\rm peak}$), and IR luminosity (\logirlum). 
We compute the mass-weighted dust temperature following \cite{Draine_Li_2007} and IR luminosity by integrating the dust SED over 3-1000$\rm \mu m$. 
The off-diagonal panels show the joint distributions and covariances between these parameters. 
Panel (d) shows the inferred SFHs. 
The solid lines show the 50th percentile of the posteriors in each time bin, and the shaded regions indicate $1\,\sigma$.

We find that the two SPS models yield substantially different inferred physical properties. 
The best-fit parameters inferred with each model are listed in Table~\ref{tab:inferred_params}.
With the additional longer-wavelength constraints from the new MIRI and NOEMA observations, the Default models clearly favour the solution with lower IR luminosity (\logirlum$=11.6^{+0.1}_{-0.2}$), shown by the red lines and contours in Fig.~\ref{fig:allphot_allmodels}. 
This solution corresponds to a dust-poor ($\tau_{\rm old}=0.16^{+0.05}_{-0.05}$), old ($\rm age_{light}=0.65^{+0.04}_{-0.04}\,\rm Gyr$), and metal-poor ($\log(Z/Z_\odot)=-0.75^{+0.03}_{-0.02}$) stellar population. 
This solution is consistent with the fiducial solution obtained in the original analysis in \cite{deGraaff2025_z4.9}.

In contrast, adopting \alphamc\ models (with \afe\ as a free parameter) results in significantly different inferred stellar and dust properties (blue lines and contours in Fig.~\ref{fig:allphot_allmodels}). 
First, the system appears to be $\alpha$-enhanced, with \afe$\approx0.4$. 
The relationship between \feh, \afe, and total metallicity [Z/H] in \alphamc\ is as follow \citep{Park2025}: [Z/H] = \feh + $A$\afe, where $A\approx0.77$ for \afe$=+0.4$.
Combining the inferred \feh\ and \afe\ therefore yields a near-solar total metallicity, $\log(Z/Z_\odot)\approx\rm [Z/H] = -0.11$
The \alphamc\ models also favour a substantially younger stellar population with $\rm age_{light}=0.26^{+0.04}_{-0.03}\,\rm Gyr$.
The corresponding SFH (Panel d) indicates a recent burst of star formation approximately 300 Myr before observation, followed by rapid quenching. 
This solution also has moderately higher dust attenuation ($\tau_{\rm old}=0.31^{+0.05}_{-0.05}$) and a higher IR luminosity (\logirlum$=12.0^{+0.1}_{-0.1}$).

\subsection{The choice of SPS models}
\begin{figure*}
    \centering
    \includegraphics[width=\textwidth]{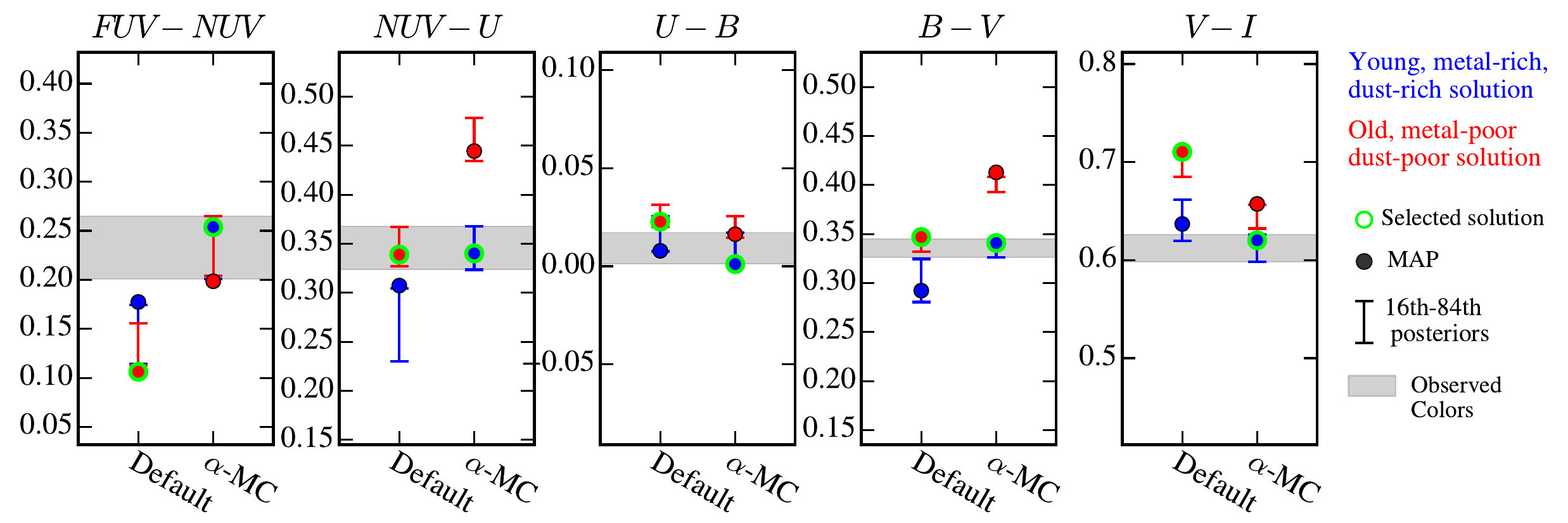}
    \caption{Predicted rest-frame UV-to-IR broadband colours for the two distinct solutions (blue: young and metal-rich solution, and red: old and metal-poor solution) under the default and \alphamc\ models. 
    The predicted colours based on MAP values of the young and old solutions are shown as circles, with error bars representing the 16th-84th percentile ranges derived from random posterior samples.  
    The grey horizontal bands indicate the observed rest-frame UV-to-IR colours covering 16th-84th percentiles. 
    Among the two solutions (blue and red), the preferred solution for each model is highlighted with green outlines. 
    We find that the two SPS models have systematically different colour predictions, especially for the rest-frame UV colours (e.g., $FUV-NUV$ and $NUV-U$), with differences up to $\sim0.1$ mag. 
    The differences in the underlying SED shapes, particularly in rest-UV, play a critical role in determining the inferred stellar population properties: The default models favour the old, metal-poor solution (red), whereas when \alphamc\ models are used, the young and metal-rich solution (blue) provides a better match to the observed colours.} 
    \label{fig:color_predictions}
\end{figure*}

The SED fitting results show that different SPS models yield distinct inferred solutions: the default models favour a metal-poor, old, and dust-poor scenario, while \alphamc\ models suggest an $\alpha$-enhanced (\afe$\approx0.4$), metal-rich, young stellar population with higher dust attenuation. 
Here, we explore why the two models yield different solutions and discuss the impact of the choice of SPS models. 
Using the best-fit model parameters of the two solutions (the young solution obtained by \alphamc\ and the old solution obtained by the default models), we compute the rest-frame UV-to-IR broadband colours for both solutions predicted by the two models.

Figure~\ref{fig:color_predictions} presents the predicted broadband colours for the two distinct solutions under the SPS models shown along the x-axis (the default and \alphamc\ models). 
The young, metal-rich, higher dust attenuation solution is colour-coded as blue, while the old, metal-poor, and dust-poor solution is shown in red. 
The predicted colours based on MAP values are shown as circles, with error bars representing the 16th-84th percentile ranges derived from random posterior samples.
The grey horizontal bands indicate the observed rest-frame UV-to-IR colours covering 16th-84th percentiles. 
Among the two solutions (blue and red), the preferred solution for each model is highlighted with green outlines.

We find that the two SPS models yield systematically different colour predictions, especially for the rest-frame UV colours (e.g., $FUV-NUV$ and $NUV-U$).
Regardless of the solutions, the two models show rest-frame UV colour differences up to $\sim0.1$ mag.
As a result, the preferred solution depends on the adopted SPS model: when the \alphamc\ models are used, the young and metal-rich solution provides a better match to the observed colours, whereas the default models favour the old, metal-poor solution. 
This demonstrates that the differences in the underlying SED shapes, particularly in rest-UV, play a critical role in determining the inferred stellar population properties.

One thing to note is that the MILES spectral library \citep{Sanchez-Blazquez2006}, adopted in the default models (MIST v1.2 + MILES), covers only the rest-frame optical range (3525-7500\text{\AA} with FWHM=2.3\text{\AA} spectral resolution). 
In FSPS \citep{Conroy2009_FSPS}, the solar-scaled C3K v1.3 spectral library \citep{Cargile2020_C3K} is therefore stitched to MILES outside the rest-optical wavelengths to ensure the full wavelength coverage. 
To explore how this affects the full spectral range fitting, we refit the same observed dataset using the default models, while limiting the NIRSpec/PRISM spectrum to the rest-optical range (i.e., the wavelength range covered by the MILES library). 
In Fig.~\ref{fig:allphot_allmodels}, the yellow lines and histograms are the SED fitting results using the same default models (MIST v1.2 + MILES), but fitting only for the rest-optical range ($0.3-0.7\,\rm \mu m$) of the NIRSpec/PRISM data.

Interestingly, with the default models, fitting only the rest-optical part of the PRISM spectrum (the native wavelength coverage of the MILES library) favours a solution comparable to that obtained with the \alphamc\ models: a young ($\rm age_{\rm light}\approx0.3\,Gyr$), metal-rich ($\log(Z/Z_\odot)\approx0.0$) stellar population with more dust attenuation ($\tau_{\rm old}\approx0.3$) and high IR luminosity (\logirlum$\approx11.9$).    
This suggests that the old solution inferred by the default SPS models is primarily driven by fitting the rest-UV continuum with inconsistent spectral models. 
Performing full wavelength-range spectrum fitting with such inconsistent spectral models -- where different spectral libraries (MILES and C3K) are stitched together across the wavelength range -- can lead to incorrect inferred properties.
This highlights the importance of using the SPS models that self-consistently cover the full wavelength range when inferring physical properties from full-spectrum fitting.

\section{Discussion}
\label{sec:discussion}

\subsection{$\alpha$-enhancement}

The \alphamc\ models (with \feh\ and \afe\ as free parameters) infer \feh$=-0.41^{+0.12}_{-0.13}$ and \afe$=+0.38^{+0.10}_{-0.10}$ (blue histograms in Fig.~\ref{fig:allphot_allmodels}), corresponding to a total metallicty of $\log(Z/Z_\odot)=-0.12^{+0.10}_{-0.10}$.
Note that \alphamc\ models adopt the solar abundance ($Z_\odot$) from \cite{GS98}.   
The inferred $\alpha$-enhancement is qualitatively consistent with recent studies suggesting that early ($z>3$) massive quiescent galaxies are mostly $\alpha$-enhanced \citep[e.g.,][]{Hamadouche2026}, indicative of rapid formation timescales, while the degree of $\alpha$-enhancement varies and depends on the methods used \citep[e.g.,][]{Leung2026}. 
Although our full spectrum fitting infers a clear $\alpha$-enhancement, the current constraints are likely provided by the rest-UV SED shapes/colours \citep[e.g.,][]{Byrne2022, Park2025}. 
At the spectral resolution of the current NIRSpec/PRISM observations ($R\sim100$), direct metallicity measurements from individual metallicity-sensitive absorption features remain limited.  
More robust measurements of \afe\ will require higher resolution spectroscopy capable of resolving the rest-frame optical features particularly sensitive to $\alpha$-enhancement, e.g., Mg b.

\subsection{Stellar mass assembly history}
\begin{figure*}
    \centering
    \includegraphics[width=\textwidth]{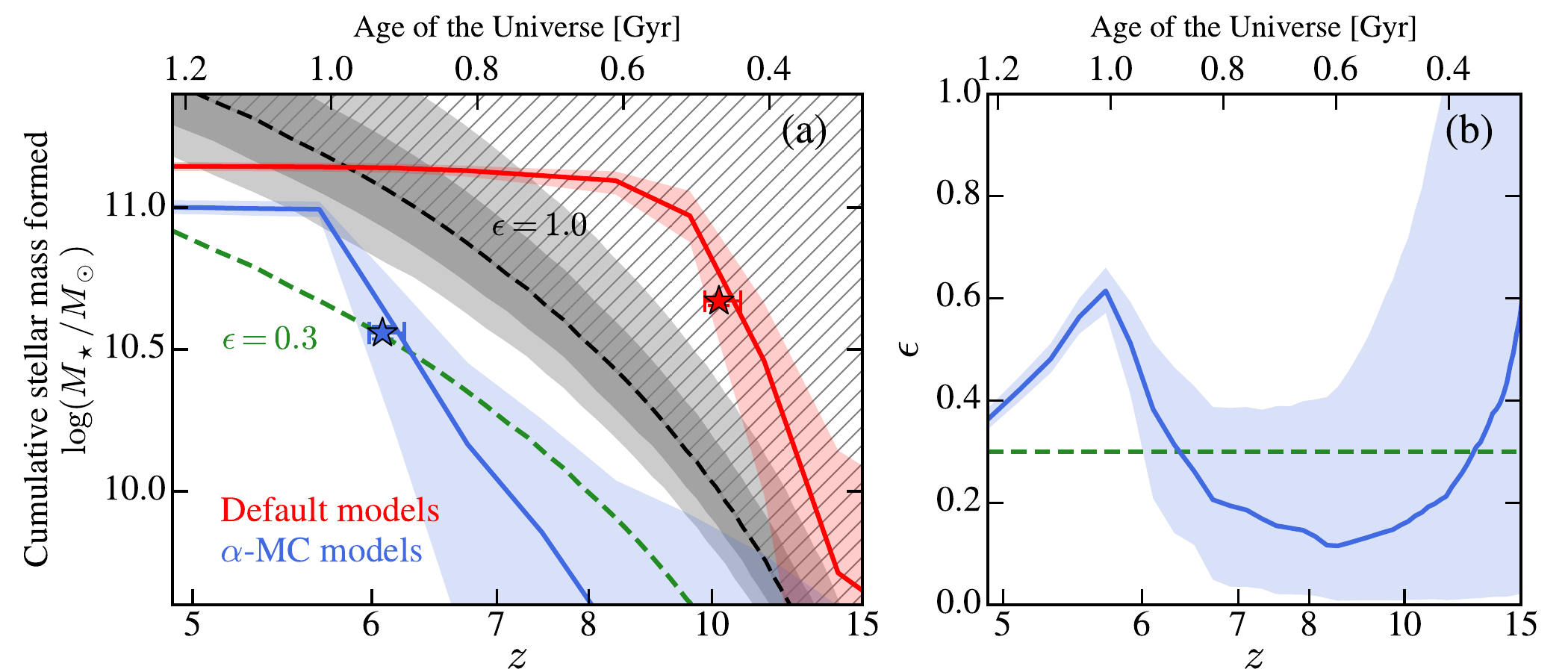}
    \caption{(a) Inferred stellar mass growth history of \galaxyname. 
    The black dashed line (with dark and light shades indicating $1\,\sigma$ and $2\,\sigma$ contours) shows the predicted stellar mass growth history of the most massive galaxy ($M_{\star,\rm max}$), i.e., maximum stellar mass possible in $\Lambda$CDM, assuming 100\% baryon-to-star conversion, i.e., $\epsilon=1.0$, computed from \citet{Lovell2023_evs}; Stellar mass growth above this line therefore requires unphyscially high ($>100\%$) baryon-to-star conversion efficiencies (hatched region). 
    The solid lines show the inferred stellar mass growth history of \galaxyname\ using the default (red line) and \alphamc\ (blue line) models. The shaded regions indicate $1\,\sigma$ uncertainties. 
    The stars indicate the formation redshift $z_{\rm form}$ (when half of the final stellar mass is formed), with the error bars representing the 16th to 84th percentile ranges. The stellar mass growth history inferred using the default models requires a baryon-to-star conversion efficiency ($\epsilon$) exceeding 100\%. When \alphamc\ models are adopted, the formation history can be explained with an overall baryon-to-star conversion efficiency of $\epsilon\approx0.3$ (green dashed line).
    (b) Baryon-to-star conversion efficiency ($\epsilon$) as a function of time, based on the SFH (blue line in Panel a) inferred from the \alphamc\ models using non-parametric SFH and continuity SFH prior. The green dashed line indicates $\epsilon=0.3$. The inferred efficiency remains relatively modest ($\epsilon\sim0.2$) during the early stage of the assembly but rises rapidly to $\epsilon\sim0.6$ at the peak of the starburst at $z\sim6$. Although substantially lower than the $>100\%$ efficiency inferred from the old solution, the inferred efficiency is still remarkably high, implying that the progenitor of \galaxyname\ (with \logmstar$\approx10.5$ and peak SFR$\approx600\,M_\odot/ \rm yr$ at $z\sim6$) would have been among the most extreme star-forming systems at this epoch.} 
    \label{fig:mgrowth}
\end{figure*}

The $\Lambda$CDM model can place an upper limit on the stellar mass ($M_{\star, \rm max}$) of the most massive galaxy at a given redshift and volume, assuming that it resides in the most massive halo at the corresponding epoch ($M_{\rm halo, max}$). 
The relationship between $M_{\star,\rm max}$ and $M_{\rm halo,max}$ can be expressed as follows: $M_{\star,\rm max} = \epsilon\,f_b\,M_{\rm halo,max}$, where $\epsilon$ is a baryon-to-star conversion efficiency and $f_b$ is cosmic baryon fraction. 
While the cosmic baryon fraction is well constrained to be $f_b=15.6\%$ \citep{Planck2020}, the baryon-to-star conversion efficiency remains highly uncertain, especially at high redshifts, where it is extremely challenging to observationally constrain.

There is growing evidence that star formation in high-redshift galaxies tends to be bursty, resulting in higher baryon-to-star conversion efficiency than local galaxies \citep[e.g.,][]{Xiao2024, Shuntov2025, Lapasia2026}.  
However, as discussed in the Introduction, some massive quiescent galaxies at $z>3$ in the literature appear to require unphysically high baryon-to-star conversion efficiency exceeding 100\% \citep[e.g.,][]{Glazebrook2024, Carnall2024excels, Weibel2025_z7.3, Turner2025, McConachie2026}. 
Among the two viable SFHs presented in the original analysis of \cite{deGraaff2025_z4.9}, the fiducial solution for \galaxyname\ also implies an impossibly high (>100\%) baryon-to-star conversion efficiency. 
Here we explore whether the inferred SFH of \galaxyname\ using the new self-consistent \alphamc\ models can be explained with a more reasonable baryon-to-star conversion efficiency, consistent with our understanding of galaxy formation in the $\Lambda$-CDM paradigm.

Fig.~\ref{fig:mgrowth} (a) shows the inferred stellar mass growth history of \galaxyname, inferred using the default (red line) and \alphamc\ (blue line) models.
The shaded regions show $1\,\rm \sigma$ uncertainties.   
The stars indicate the formation redshift $z_{\rm form}$, defined as the redshift at which half of the final stellar mass is formed. 
The error bars represent the 16th to 84th percentile range of $z_{\rm form}$. 
The black dashed line (with dark and light shades indicating $1\,\sigma$ and $2\,\sigma$ contours) shows the predicted stellar mass growth history of the most massive galaxy ($M_{\star,\rm max}$) assuming 100\% baryon-to-star conversion efficiency. 
Stellar mass growth histories above this line (hatched region) would therefore require unphysically high baryon-to-star conversion efficiencies exceeding 100\%.
The predicted mass of the most massive galaxy is computed from \cite{Lovell2023_evs}, which uses extreme value statistics (EVS) to provide the confidence regions in the mass-redshift plane, independent of the selection function. 
We assume a survey area of $A=150\,\rm arcmin^2$, comparable to those of \cite{Carnall2024excels} and \cite{Valentino2023ApJ}, to compute the predicted mass of the most massive galaxy.

The SFH inferred using \alphamc\ (blue line) suggests that the formation history of \galaxyname\ can be explained with a more physically plausible baryon-to-star conversion efficiency. 
The stellar population inferred using \alphamc\ is younger by $\approx0.3\,\rm Gyr$, corresponding to a much lower formation redshift of $z_{\rm form}=6.1^{+0.2}_{-0.1}$ (blue star), compared to $z_{\rm form}=10.2^{+0.3}_{-0.3}$ (red star) inferred using the default models. 
The final stellar mass is also lower by $\approx0.2$ dex. 
With younger age and lower stellar mass, the formation history of \galaxyname\ implies an overall baryon-to-star conversion efficiency of $\epsilon\approx0.3$\footnote{Time-averaged baryon-to-star conversion efficiency. It also corresponds to the baryon-to-star conversion efficiency at the formation redshift (blue star). If averaged only for $z<10$ where SFH is more robustly constrained, the averaged efficiency is $\epsilon=0.26$.} (green dashed line). 
The inferred efficiency of $\epsilon\approx0.3$ is consistent with the values inferred for many high-$z$ galaxies \citep[e.g.,][]{Xiao2024, Shuntov2025}.

While the overall efficiency is $\epsilon\approx0.3$, the galaxy may still require a remarkably high conversion efficiency during its main growth phase.
Fig.~\ref{fig:mgrowth} (b) shows the baryon-to-star conversion efficiency ($\epsilon$), based on the assembly history inferred using \alphamc\ models, as a function of redshift. 
We note that the detailed shape of the inferred SFH, and consequently, the evolution of $\epsilon$, depends strongly on the adopted SFH parametrization and prior; therefore, the overall baryon-to-star conversion efficiency of $\epsilon\approx0.3$ is more robustly constrained than its detailed evolution.
The original analysis of \cite{deGraaff2025_z4.9} has explored different SFH parametrizations and priors and found broadly consistent formation histories. 
As the primary focus of this work is to investigate the effect of adopting different SPS models, we adopt the same non-parametric SFH and continuity prior as in the original analysis.
The SFH inferred under this setup suggests that the efficiency remains relatively modest ($\epsilon\sim20\%$) during the early stage of the assembly, but it increases rapidly and reaches $\epsilon\sim60\%$ at the peak of the starburst at $z\sim6$. 
Although this is substantially lower than the physically implausible efficiency exceeding $\epsilon>100\%$ required by the old solution inferred from the default models (red line in Panel a), the peak efficiency of $\approx60\%$ is still remarkably high. 
This suggests that, if observed during its peak growth phase, its progenitor (with a stellar mass of \logmstar$\approx10.5$) would have appeared as one of the most extremely star-forming systems (with peak SFR$\approx600\,M_\odot/ \rm yr$) at $z\sim6$ \citep[e.g.,][]{Riechers2013, Strandet2017, Zavala2018}. 

\vspace{0.5cm}
% implication for other maximally old galaxies
Our results may have broader implications for the assembly histories of other ``maximally old'' galaxies found at $z>3$, which appear to require baryon-to-star conversion efficiencies exceeding $100\%$.  
Using the self-consistent \alphamc\ models, we infer an $\alpha$-enhanced (\afe$\sim0.4$) and younger stellar population for \galaxyname.
Although the \afe\ constraint in our analysis is primarily driven by rest-UV colours \citep[e.g.,][]{Byrne2022} due to low PRISM resolution ($R\sim100$), many maximally old galaxies in the literature (typically with $z_{\rm form}\gtrsim10$)  have been observed with medium- or high-resolution NIRSpec spectroscopy \citep[e.g.,][see also JWST Cycle 3, GO 5069]{Carnall2024excels, Park2024, McConachie2026}, resolving the key rest-optical features (including Mg b, Ca, and Fe lines), from which \afe\ can be constrained much more robustly. 
Indeed, \cite{Carnall2024excels} have measured $\rm [Mg/Fe]=0.42$ for ZF-UDS-7329, a maximally old galaxy at $z=3.2$ \citep{Glazebrook2024}.
If these maximally-old systems are similarly $\alpha$-enhanced, incorporating self-consistent $\alpha$-enhanced stellar population models could potentially lead to systematically younger inferred stellar ages and lower stellar masses \citep[e.g.,][]{Park2025}, thereby substantially reducing the required baryon-to-star conversion efficiencies.

Several other factors have also been proposed to reconcile the apparent extreme baryon-to-star conversion efficiency either by increasing the predicted halo mass or by reducing the inferred stellar ages or masses. 
First, many massive quiescent galaxies at high redshifts appear to reside in overdense environments \citep[e.g.,][]{McConachie2022, Valentino2023ApJ, Tanaka2024, McConachie2025_rubies}. 
Incorporating galaxy environments into the EVS calculation, \cite{Jespersen2025} have shown that conditioning on overdensities increases the maximum halo and stellar masses, substantially reducing the required efficiency. 
On the observational side, the choice of SFH priors adopted in the SED models can also affect the inferred stellar ages. 
\cite{Turner2025} demonstrated that adopting a physically motivated rising SFH prior, motivated by dark matter accretion histories, leads to systematically younger inferred stellar ages for these maximally-old galaxies. 
They further found that accounting for stellar mass growth through major mergers \citep[e.g.,][]{Ito2025_merger} can significantly reduce the required baryon-to-star conversion efficiency down to $\epsilon\approx0.5$. 
Finally, using JWST medium-band photometry, \cite{Cutler2026} have found that several maximally old galaxies exhibit negative colour gradients \citep[see also][]{Turner2025}. 
Because NIRSpec slit spectroscopy preferentially samples the redder central regions, unresolved analysis can slightly overestimate the total stellar mass of these massive quiescent galaxies with strong colour gradients. 
Accounting for the colour gradients can reduce the inferred stellar masses by $\sim0.1$ dex, which can help alleviate the apparent tension.

% IMF
In this work, we adopt the same Chabrier IMF as in the original analysis of \cite{deGraaff2025_z4.9} to assess more clearly the impact of the additional IR constraints and the $\alpha$-enhanced stellar population models. 
However, the IMF remains one of the significant sources of systematic uncertainties in stellar population modelling that has not been explored in this work. 
Several studies have shown that the IMF has significant impacts on the inferred stellar masses \citep[e.g.,][]{Wang2024_quantifying}. 
There has been growing evidence that some high-redshift galaxies tend to have top-heavy IMFs \citep[e.g.,][]{Katz2022, Steinhardt2023, Yung2024, Cameron2024}. 
On the other hand, observations of massive early-type galaxies in the local Universe have suggested relatively bottom-heavy IMFs \citep[e.g.,][]{Treu2010, Conroy2012, Conroy2017}.
More recently, \cite{Cheng2026} have shown that massive quiescent galaxies at $z\sim0.7$ appear to have bottom-heavy IMFs, which increases the inferred stellar masses of their oldest samples by a factor of $\sim3-4$. 
If \galaxyname\ also has a similarly bottom-heavy IMF, the true stellar mass would likewise be higher than inferred in this work, increasing the required baryon-to-star conversion efficiency.  
Future work providing robust measurements of the IMF of this galaxy and other maximally-old galaxies will therefore be essential in assessing the true severity of the apparent tension with galaxy formation models.

In summary, our results suggest that uncertainties associated with stellar population models should be considered alongside these other factors when interpreting the formation histories of these maximally old galaxies. 
In particular, our work demonstrates the importance of using self-consistent SPS models with non-solar \afe\ when inferring the physical properties of high-redshift massive quiescent galaxies which are likely to be $\alpha$-enhanced. 
Together with the effects of overdense environments, merger-driven assembly, assumptions about the SFH prior, and spatially resolved stellar population gradients, such improved SPS models could provide an additional pathway toward reconciling the assembly histories of these galaxies with our current understanding of galaxy formation. 
Our results raise the possibility that at least some of the galaxies currently inferred as ``maximally old'' with solar-scaled models may instead host younger, less massive, and $\alpha$-enhanced stellar populations with much more physically plausible assembly histories.

\subsection{Dust temperature and IR luminosity}

\begin{figure}
    \centering
    \includegraphics[width=\columnwidth]{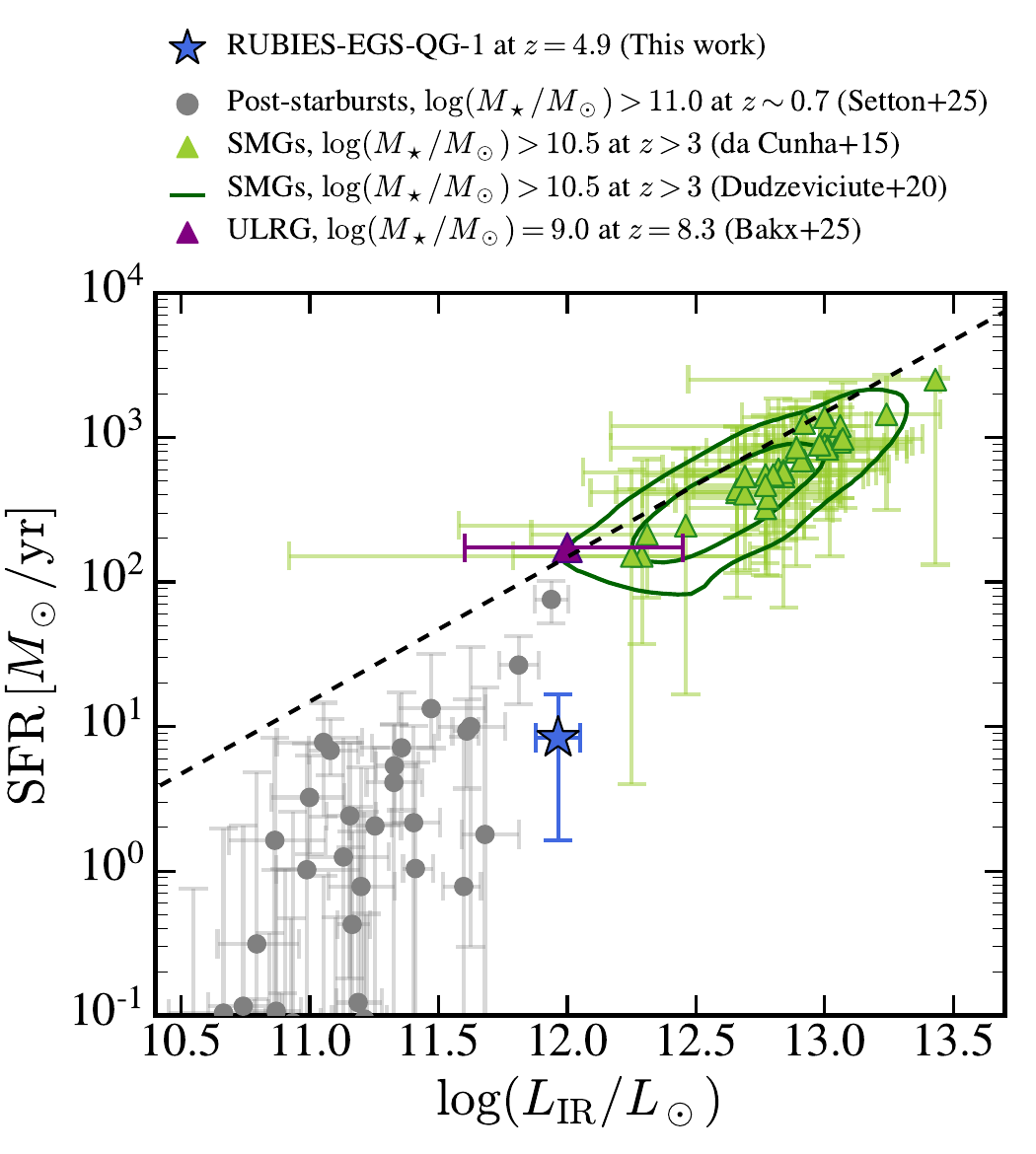}
    \caption{Inferred SFR vs. IR luminosity of \galaxyname\ (blue star, this work), compared to massive ($>10^{11}\,M_\odot$) post-starburst population at $z=0.7$ (gray circles; \citet{Setton2025}), massive ($>10^{10.5}\,M_\odot$) SMG populations at $z>3$ (green triangles; \citealt{daCunha2015} and green contours; \citealt{dudzeviciute2020}), and an ULIRG at $z=8.3$ \citep{Bakx2025}, purple triangle). 
    All the error bars represent the 16th to 84th percentiles. 
    The black dashed line indicates the $L_{\rm IR}$-SFR relation from \citet{Kennicutt2012}. 
    We find that \galaxyname\ shows a high IR luminosity despite its quiescent nature, placing it between the quiescent and SMG populations. 
    Given that \galaxyname\ has a recent starburst $\sim300\,\rm Myr$ ago, followed by rapid quenching, this suggests that the high inferred IR luminosity of this galaxy is likely powered by dust heating by the stars formed during the recent starburst 300 Myr ago.
    This might imply that \galaxyname\ may be in a transitional phase between an IR-bright starburst and a fully quiescent and IR-faint system.}
    \label{fig:dust_emission}
\end{figure}

Despite its quiescent nature with relatively low dust attenuation ($A_V\approx0.3\,\rm mag$), our best-fit solution inferred from the \alphamc\ models predicts that \galaxyname\ has an IR luminosity of \logirlum$\sim12.0$ and a dust temperature of $T_{\rm dust}\sim40\,\rm K$ with the peak wavelength of the dust SED at $\lambda_{\rm peak}\sim40\,\rm \mu m$. 
Note that we use the mass-weighted dust temperature $T_{\rm dust}$, following \cite{Draine_Li_2007}.
We caution that the current observations do not directly constrain the far-IR dust emission, except that the NOEMA upper limit rules out cold dust temperatures at high $L_{\rm IR}$. 
The inferred $T_{\rm dust}$ and IR luminosity are highly model-dependent predictions of the SED fitting, primarily constrained through the assumed energy balance. 
In this section, we therefore explore the physical implications of the best-fit solution inferred from the \alphamc\ model, while emphasizing that direct far-IR observations are required to robustly determine the dust properties.

Fig.~\ref{fig:dust_emission} shows the inferred IR luminosity and SFR of \galaxyname, where the SFR is averaged over $100\,\rm Myr$. 
The black dashed line indicates the $L_{\rm IR}$-SFR relation for constant SFHs from \cite{Kennicutt2012}. 
We compare the inferred IR luminosity of \galaxyname\ to massive sub-mm galaxy (SMG) populations at $z>3$ \citep[][green triangles and contours]{daCunha2015, dudzeviciute2020} and a ULIRG at $z=8.3$ \citep[][purple triangle]{Bakx2025}. 
To provide a more appropriate comparison to \galaxyname, we restrict the SMG samples of \cite{daCunha2015} and \cite{dudzeviciute2020} to only the massive ($>10^{10.5}\,M_\odot$) and high-$z$ ($z>3$) sources, resulting in 35 and 220 galaxies, respectively. 
Finally we also include massive ($>10^{11}\,M_\odot$) quiescent/post-starburst galaxies at lower redshifts, $z\sim0.7$ \citep[][grey circles]{Setton2025}. 
While the high-redshift ULIRG and SMG populations follow the canonical $L_{\rm IR}$-SFR relation, quiescent/post-starburst galaxies at lower redshifts remain IR-luminous despite their low SFRs. 
Similarly, \galaxyname\ is also predicted to remain IR-luminous, placing it between the quiescent and SMG populations.

If the high inferred IR luminosity ($L_{\rm IR}\sim10^{12}\,L_\odot$) is confirmed, it could be powered by dust heating by the stars formed during the recent starburst $\approx300\,\rm Myr$ ago.
Using hydrodynamic simulations, \cite{Hayward2014} have shown that the IR luminosity of post-starburst systems remains elevated even after rapid quenching, as the stars formed during the preceding starburst can remain dust-obscured and continue to heat the dust. 
They found that IR luminosity can overestimate the instantaneous SFR of a post-starburst system by up to two orders of magnitude \citep[see also][]{Utomo2014}. 
The SFH inferred from \alphamc\ models has shown that \galaxyname\ experienced a major starburst $\rm \sim300\,\rm Myr$ before observation, followed by rapid quenching. 
At the time of observation, the intermediate-age ($\rm \sim300\,\rm Myr$) stellar populations can still contribute to the UV-optical radiation field \citep[e.g.,][]{Cortese2008,
DeLooze2014, Leja2019_older}, heating the dust even after the SFR has substantially declined \citep[e.g.,][]{French2015, Wild2025, Lorenzon2025}. 
This suggests that, in the case of post-starburst systems, the IR luminosity may represent a delayed signature of the recent starburst episode, rather than an ongoing obscured starburst as in LIRGs/ULIRGs.

The compact structure of \galaxyname\ may further contribute to its warm dust temperature. 
Several observations have reported hot dust temperatures for high-redshift dusty star-forming/SMGs and LIRGs/ULIRGs (e.g., \citealt{Sommovigo2022a, Sommovigo2022b, Cooper2022, Cooper2025, Bakx2025,  Casey2026}, see also the models of \citealt{Imara2018}). 
Interestingly, the preferred SED solution predicts a similarly warm dust temperature for \galaxyname\ despite having relatively low dust attenuation ($A_V\sim0.3$).  
One possible explanation is its exceptionally compact structure. 
Theoretical studies have also shown that dust temperature depends highly on the surface star formation rate density $\Sigma_{\rm SFR}$ \citep[e.g.,][]{Parente2026}; more compact galaxies have higher radiation energy densities, leading to more efficient dust heating.
\galaxyname\ hosts an exceptionally dense, A-type-dominated stellar population with $R_{\rm e}\approx0.5\,\rm kpc$ and $M_\star\sim10^{11}\,M_\odot$. 
The compact configuration of intense A-star radiation could substantially increase the energy density experienced by the dust, resulting in such a high dust temperature despite low current SFR.

Several other heating sources may have also contributed to the warm dust temperature and IR emission observed in \galaxyname. 
First, residual obscured star formation may contribute to the IR luminosity. However, the combination of NIRCam, MIRI, and NOEMA observations incorporated into the SED fitting places significant constraints on the dust-obscured energy budget, limiting the amount of hidden obscured star formation. 
The AGN activity may also contribute to dust heating, particularly at mid-IR wavelengths \citep[e.g.,][]{Mullaney2011}, although AGN-heated dust can, in some cases, contribute to the far-IR emission as well \citep[e.g.,][]{Symeonidis2016}. 
The detection of emission lines, such as [O III], [N II], and [S II], and the [N II]/$\rm H\alpha$ ratio is consistent with the presence of AGN activity \citep{deGraaff2025_z4.9}, although we cannot completely rule out the possibility of a shock-ionized origin \citep[e.g.,][]{Kewley2006}. 
Finally, the cosmic microwave background also sets a temperature floor of $T_{\rm CMB}\approx17\,\rm K$ at $z=4.9$ \citep[e.g.,][]{daCunha2013}. 
While this is substantially lower than the inferred dust temperature, its contribution to the dust energy budget is not entirely negligible.

Finally, we emphasise again that the inferred dust temperature and IR luminosity remain subject to the uncertainties associated with the stellar population models. 
In our analysis, the IR luminosity is primarily constrained through energy balance, where all dust-absorbed starlight in the rest-UV-optical is re-emitted in the IR, together with the MIRI and NOEMA constraints.   
Therefore, the inferred dust properties depend strongly on the predicted intrinsic UV continuum, which is sensitive to the uncertainties in stellar population models \citep[see the review of ][]{Conroy2013}, including stellar evolution prescriptions, stellar spectral templates \citep[e.g.,][]{Byrne2023}, chemical abundance patterns \citep[e.g.,][]{Byrne2022, Park2025}, and binary evolution \citep[e.g.,][]{Eldridge2017, Stanway2018}.  
Ultimately, additional observations that directly constrain the far-IR SED, particularly at frequencies between the current MIRI and NOEMA constraints, will be essential for robustly inferring the dust properties of this galaxy and testing the physical interpretation proposed in this work.

\section{Summary and Conclusion}
\label{sec:summary_conclusion}

In this paper, we re-analyse \galaxyname\ originally studied in \cite{deGraaff2025_z4.9} by incorporating 1) new constraints from MIRI and NOEMA observations and 2) new self-consistent stellar population models with non-solar \afe, \alphamc. 
These improvements enable us to reassess its SFH and determine whether its formation history can be explained with a more physically plausible baryon-to-star conversion efficiency. 
We summarise our key findings below. 

\begin{enumerate}[itemsep=0.6em]
    
    \item Incorporating additional constraints from the new MIRI and NOEMA observations, we find that the preferred solution depends strongly on the adopted stellar population models (Fig.~\ref{fig:allphot_allmodels}). 
    The default solar-scaled models continue to favour an old ($0.6\,\rm Gyr$), metal-poor ($\log(Z/Z_\odot)=-0.7$), dust-poor ($A_V=0.17$) solution. 
    In contrast, adopting the \alphamc\ models (with \afe\ as a free parameter) reveals a younger ($0.3\,\rm Gyr$), $\alpha$-enhanced (\afe$\sim0.4$) stellar population with near-solar total metallicity ($\log(Z/Z_\odot)=-0.1$) and higher dust attenuation ($A_V=0.32$).

    \item We find that the old solution inferred using the default SPS models is primarily driven by fitting the rest-UV continuum with inconsistent spectral models. 
    The two SPS models (default and \alphamc) predict systematically different rest-frame UV SED shapes, with rest-UV colour differences up to $\sim 0.1\,\rm mag$, leading to different interpretations of the same observational data (Fig.~\ref{fig:color_predictions}).
    Notably, when the SED fitting to the PRISM spectrum is restricted to the rest-frame optical wavelength range (the native coverage of the MILES library), the default models also shift toward the young solution preferred by the \alphamc\ models.
    This suggests that performing full-spectrum fitting with such inconsistent spectral models -- where different spectral libraries are stitched together across the wavelength range -- can lead to incorrect inferred properties.
    This highlights the importance of using self-consistent SPS models that cover the full wavelength range when inferring physical properties from full-spectrum fitting.

    \item The SFH inferred using the \alphamc\ models shows that \galaxyname\ experienced a major starburst $\approx300\,\rm Myr$ before observation, followed by rapid quenching. 
    This inferred formation history significantly alleviates the apparent tension with galaxy formation models, with an overall baryon-to-star conversion efficiency of $\epsilon\sim30\%$ (Fig.~\ref{fig:mgrowth}), by lowering both the inferred stellar age (by $\sim0.3\,\rm Gyr$) and mass (by $\sim0.2\,\rm dex$). 
    However, the galaxy may still require a remarkably high efficiency at the peak of the starburst.
    Thus, although the new solution inferred using \alphamc\ no longer requires the physically implausible efficiency exceeding 100\% inferred from the old solution, it still implies an extreme formation episode during its main growth phase.

    \item Despite being quiescent and relatively dust-poor ($A_V\sim0.3$), our best-fit solution inferred using the \alphamc\ models predicts a high IR luminosity of $L_{\rm IR}\approx10^{12}\,L_\odot$ and warm dust temperature of $T_{\rm dust}\approx40\,\rm K$. 
    The high inferred IR luminosity of this galaxy is likely powered by dust heating by the stars formed during the recent starburst $\approx300\,\rm Myr$ before observation. 
    The exceptionally compact structure ($R_{\rm eff}=0.5\,\rm kpc$) of this A-type-dominated stellar population may further contribute to its warm dust temperature.
    Our results suggest that massive post-starburst galaxies at high redshifts can remain IR-luminous for several hundred Myr even after quenching. This might imply that \galaxyname\ is in a transitional phase between an IR-bright starburst and a fully quiescent IR-faint system (Fig.~\ref{fig:dust_emission}) where the star formation has already been quenched, while the dust reservoir and associated IR emission have not yet fully faded.

\end{enumerate}

Our reanalysis of \galaxyname\ using the new self-consistent SPS models suggests that it is a young, IR-luminous, post-starburst galaxy, rather than a maximally-old system. 
The inferred assembly history can be reconciled with substantially lower baryon-to-star conversion efficiencies ($\sim30\%$), although still requiring a remarkably high peak efficiency of $\sim60\%$ during the starburst phase. 
Nevertheless, additional observations are needed to test this interpretation. 
Higher-resolution NIRSpec spectroscopy will be essential for robustly constraining the degree of $\alpha$-enhancement in this system.
Future observations that directly constrain the far-IR SED, particularly at wavelengths between the current MIRI and NOEMA constraints, will provide much stronger constraints on the dust properties of this system.

More broadly, our results may have important implications for other maximally old galaxies found in the literature. 
If $\alpha$-enhancement is widespread among this massive quiescent population at high redshifts, some of these galaxies currently inferred as ``maximally old'' might instead be younger and less massive, with assembly histories that can be explained with more physically plausible baryon-to-star conversion efficiencies. 
Applying physically self-consistent SPS models taking into account non-solar \afe\ to these high-$z$ massive quiescent systems will therefore be an important next step toward determining whether the apparent tension with galaxy formation models (i.e., implausibly high star formation efficiency) reflects genuinely extreme early galaxy assembly or systematic uncertainties in current stellar population models.

\section*{Data Availability}
Data directly related to this publication and the figures are available on request from the corresponding author. 
The JWST/NIRCam, MIRI, and NIRSpec data are publicly available on the DAWN JWST Archive (DJA\footnote{https://dawn-cph.github.io/dja/index.html}).

\section*{Acknowledgements}
This work is based on observations carried out under project number W24EE with the IRAM NOEMA Interferometer. IRAM is supported by INSU/CNRS (France), MPG (Germany) and IGN (Spain).
MP acknowledges support from the Kavli Institute for Cosmology, Cambridge and the Kavli Foundation, and King's College, Cambridge.
FV acknowledges support from the Independent Research Fund Denmark (DFF) under grant 3120-00043B.
AdG acknowledges support from a Clay Fellowship awarded by the Smithsonian Astrophysical Observatory. AdG acknowledges funding by the European Union (ERC, FIRST-GIANTS, 101221926). 
I.M. acknowledges the support from Grant 63667 from the John Templeton Foundation. Any opinions, findings, and conclusions or recommendations expressed in this material are those of the authors and do not necessarily reflect the views of the John Templeton Foundation.


\bibliographystyle{mnras}
\bibliography{references} % if your bibtex file is called example.bib

%\begin{thebibliography}{99}
%\end{thebibliography}

%%%%%%%%%%%%%%%%%%%%%%%%%%%%%%%%%%%%%%%%%%%%%%%%%%

%%%%%%%%%%%%%%%%% APPENDICES %%%%%%%%%%%%%%%%%%%%%

\appendix

\section{SED fitting of the original dataset with more flexible model parameters}

\begin{figure*}
    \centering
    \includegraphics[width=0.95\textwidth]{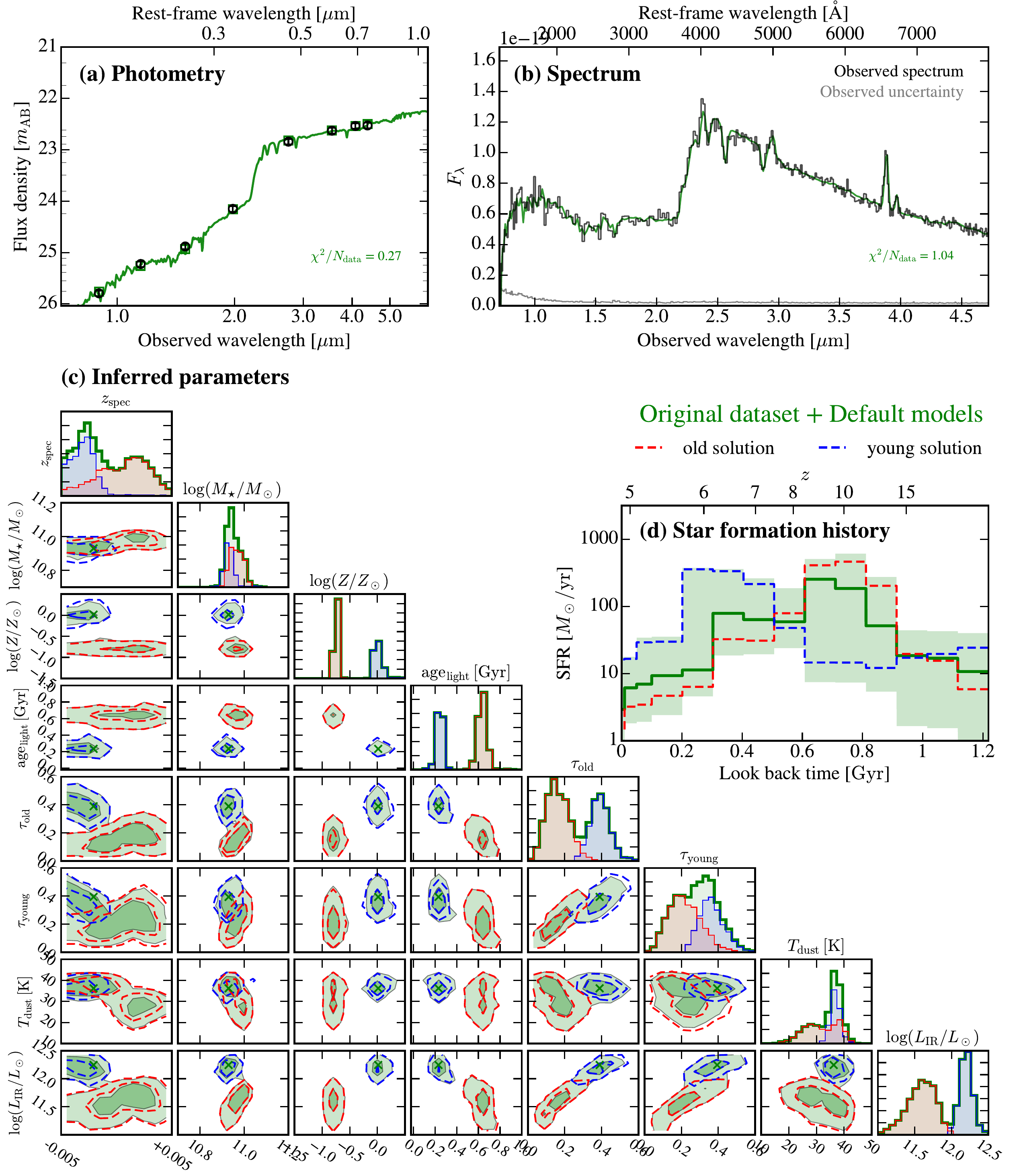}
    \caption{Prospector SED-fitting results using the original observed dataset (8-band NIRCam photometry + NIRSpec/PRISM spectrum) and default SPS models, where additional model parameters are allowed to vary. 
    (a) Fit to photometry. Observed 8-band NIRCam photometry is shown as black circles with uncertainties shown as error bars. The best-fit SED model is shown as a green line. 
    (b) Fit to the spectrum. Observed NIRSpec/PRISM data is shown as a black line, with uncertainties shown as a grey line. The best-fit spectrum is shown in green. 
    (c) Corner plot illustrating the joint distributions and covariances between model parameters. 
    The diagonal panels show the posterior distributions of each model parameter, from left to right: redshift ($z_{\rm spec}$), final (survived) stellar mass ($\log(M_\star/M_\odot)$), total metallicity ($\log(Z/Z_\odot)$), light-weighted age, $\tau_{\rm old}$, $\tau_{\rm young}$, dust temperature ($T_{\rm dust}$), and IR luminosity (\logirlum). 
    The contours in the off-diagonal panels indicate $1\,\rm \sigma$ and $2\,\rm \sigma$ of the joint posterior distributions, and the crosses indicate the MAP values.  
    (d) Nonparametric SFH inferred using the continuity SFH prior. The solid line shows the 50th percentile of the posteriors in each time bin, with the shaded region indicating $1\,\sigma$.
    We find that the original observational dataset suggests a bimodal solution for this galaxy (i.e., two degenerate solutions with similar likelihoods). 
    The posterior distributions of the two solutions are overplotted as red and blue contours, and their corresponding SFHs as red (old posteriors) and blue (young posteriors) lines. 
    The bimodal solution reflects a clear degeneracy between age, metallicity, and dust, with two distinct scenarios: an old, metal-poor, dust-poor population vs. a young, metal-rich population with higher dust attenuation.}
    \label{fig:nc_bimodal}
\end{figure*}

As a diagnostic test of the original analysis, we refit the same observation dataset (NIRCam 8-band photometry + NIRSpec/PRISM spectrum) \citep{deGraaff2025_z4.9} using the default SPS models, while allowing some model parameters to vary. 
The purpose of this re-analysis is to assess the sensitivity of the solution inferred from the original dataset to small variations in the model parameters and to characterize the degeneracies present before incorporating the new MIRI and NOEMA constraints.
In the original analysis in \cite{deGraaff2025_z4.9}, some model parameters are assumed to be fixed: $z_{\rm spec}=4.9086$, $\tau_{\rm young}=\tau_{\rm old}$, $q_{\rm PAH}=2.0$, $\gamma=0.01$, and $U_{\rm min}=1.0$. 
Here we allow the redshift to vary between [$z_{\rm spec}-0.005$, $z_{\rm spec}+0.005$] and use a clipped normal prior the ratio $\tau_{\rm young}$/$\tau_{\rm old}\sim N(\mu=1, \sigma=0.3)$, truncated at 0.0 (minimum) and 2.0 (maximum). 
See Table~\ref{tab:prospector_params} for the details of the priors used. 
Fig.~\ref{fig:nc_bimodal} presents the SED fitting results, in the same format as in Fig.~\ref{fig:allphot_allmodels}.

We find that allowing more model parameters, including redshift, to vary reveals a bimodal solution for \galaxyname\ (i.e., two degenerate solutions with similar likelihoods). 
We overplot the posterior distributions of the two solutions as red (old solution; posteriors with $\rm age_{\rm light}>0.4\,\rm Gyr$) and blue (young solution; posteriors with $\rm age_{\rm light}<0.4\,\rm Gyr$) contours in Panel (c), and their corresponding SFHs as red and blue dashed lines in Panel (d). 
The bimodal solution reflects a clear degeneracy between age, metallicity, and dust, with two distinct scenarios: (1) an old ($\rm age_{\rm light}\approx0.6\,\rm Gyr$), metal-poor ($\log(Z/Z_\odot)\approx-0.7$), dust-poor ($\tau_{\rm old}\approx0.1$) solution, and (2) a young ($\rm age_{\rm light}\approx0.3\,\rm Gyr$), metal-rich ($\log(Z/Z_\odot)\approx+0.0$) solution with higher dust attenuation ($\tau_{\rm old}\approx0.4$).
This re-analysis reveals the shallow and degenerate nature of the solution space constrained by the original observational dataset, such that small variations in the model parameters (e.g., redshift) can easily shift the solution to the alternative mode. 
The two solutions also predict distinct IR luminosities, illustrating why the additional MIRI and NOEMA constraints incorporated in the main analysis provide useful information for distinguishing between these degenerate solutions. 
As shown in Section~\ref{sec:fitting_new_data_new_models}, however, the preferred solution ultimately also depends on the adopted SPS models.

%%%%%%%%%%%%%%%%%%%%%%%%%%%%%%%%%%%%%%%%%%%%%%%%%%

% Don't change these lines
\bsp	% typesetting comment
\label{lastpage}
\end{document}